\documentclass[letterpaper]{article} 
\usepackage{aaai25}  
\usepackage{times}  
\usepackage{helvet}  
\usepackage{courier}  
\usepackage[hyphens]{url}  
\usepackage{graphicx} 
\usepackage{natbib}  
\usepackage{caption} 
\usepackage{algorithm}
\usepackage{algorithmic}
\usepackage{amssymb} 
\usepackage{amsmath}
\usepackage{bm}
\usepackage{pdfpages}
\usepackage{booktabs} 
\usepackage{multirow} 
\usepackage{comment}
\usepackage{tabularx} 
\usepackage{adjustbox}
\usepackage{subcaption}
\usepackage{newfloat}
\usepackage{listings}
\DeclareCaptionStyle{ruled}{labelfont=normalfont,labelsep=colon,strut=off} 
\floatstyle{ruled}
\newfloat{listing}{tb}{lst}{}
\floatname{listing}{Listing}
\title{CSL-L2M: Controllable Song-Level Lyric-to-Melody Generation Based on Conditional Transformer with Fine-Grained Lyric and Musical Controls}
\author{
    Li Chai,
    Donglin Wang\textsuperscript{}\thanks{Corresponding author.}
}
\affiliations{
    Westlake University\\
    \{chaili, wangdonglin\}@westlake.edu.cn
}

\usepackage{bibentry}
\begin{document}

\maketitle

\begin{abstract}

Lyric-to-melody generation is a highly challenging task in the field of AI music generation. Due to the difficulty of learning strict yet weak correlations between lyrics and melodies, previous methods have suffered from weak controllability, low-quality and poorly structured generation. To address these challenges, we propose CSL-L2M, a controllable song-level lyric-to-melody generation method based on an in-attention Transformer decoder with fine-grained lyric and musical controls, which is able to generate full-song melodies matched with the given lyrics and user-specified musical attributes. Specifically, we first introduce REMI-Aligned, a novel music representation that incorporates strict syllable- and sentence-level alignments between lyrics and melodies, facilitating precise alignment modeling. Subsequently, sentence-level semantic lyric embeddings independently extracted from a sentence-wise Transformer encoder are combined with word-level part-of-speech embeddings and syllable-level tone embeddings as fine-grained controls to enhance the controllability of lyrics over melody generation. Then we introduce human-labeled musical tags, sentence-level statistical musical attributes, and learned musical features extracted from a pre-trained VQ-VAE as coarse-grained, fine-grained and high-fidelity controls, respectively, to the generation process, thereby enabling user control over melody generation. Finally, an in-attention Transformer decoder technique is leveraged to exert fine-grained control over the full-song melody generation with the aforementioned lyric and musical conditions. Experimental results demonstrate that our proposed CSL-L2M outperforms the state-of-the-art models, generating melodies with higher quality, better controllability and enhanced structure.

\end{abstract}

\begin{links}
\link{Demos}{https://lichaiustc.github.io/CSL-L2M/}
\link{Code}{https://github.com/LiChaiUSTC/CSL-L2M}
\end{links}

\section{Introduction}
Deep learning techniques have been increasingly applied to various music generation tasks \cite{duan2024semantic, hahn2023interpretable, yu2021conditional, tian2023unsupervised}. 
Lyric-to-melody generation, one of the most essential and common tasks in songwriting, has attracted growing interest from both academia and industry. A high-quality lyric-to-melody generation is required to generate melodies not only following good musical patterns but also aligning with the given lyrics. Due to the scarcity of paired lyric-melody data with alignment information and the difficulty of learning the strict but weak correlations between lyrics and melodies, this task remains under-explored.

Many deep learning methods have been explored for lyric-to-melody generation. 
A sequence-to-sequence based melody composition model is proposed in \cite{bao2019neural} , which is the first work to use an end-to-end network model to generate melodies from lyrics. Subsequently, Yu \cite{yu2021conditional} proposes a conditional LSTM-GAN generative model for melody generation from lyrics. In \cite{srivastava2022melody}, a novel architecture, three branch conditional LSTM-GAN is proposed to further improve generation quality.
However, the direct mapping from lyrics to melodies is difficult to learn because they are weakly correlated (e.g., a melody can correspond to many different lyrics and vise versa.). Accordingly, these end-to-end generation methods suffer from low generation quality due to the limited available parallel lyric-melody data. To this end, an unsupervised method is proposed in \cite{sheng2021songmass}, which performs self-supervised masked sequence to sequence pre-training on large amount of unpaired lyric and melody data. In addition, a two-stage generation method with music template is proposed in \cite{ju2021telemelody}, which is data efficient and addresses the issues of limited paired data to some extent. With the tremendous success of large language models (LLMs) \cite{touvron2023llama}, more recently, the work in \cite{ding2024songcomposer} attempts to leverage the capability of LLMs to model the lyric-melody relationship.

Currently, the lyric-to-melody generation methods including those mentioned above focus on generating short melodies from lyrics typically consisting of one sentence or a few sentences, where a full-song melody is usually composed by simply concatenating these sentence-level melodies resulting in incoherent musical structure without both repetition patterns and distinguishable verse-chorus structure.
In addition, controllability is a crucial aspect of the lyric-to-melody generation task, which allows users to interact with the generation process to create their expected melodies. Nevertheless, only a few lyric-to-melody works have explored the controllability. In \cite{ju2021telemelody}, the generated melodies can be controlled by adjusting the musical elements in music templates including tonality and chord progression. A reference style embedding technique is proposed in \cite{zhang2023controllable}  to achieve the control over the style of generated melodies. The research of \cite{duan2022interpretable} enables users to interact with the generation process and recreate music by selecting from recommended musical attributes.
However, these works only provide a few coarse-grained musical attribute controls. 
One more thing, since one syllable may correspond to one or more notes, the alignment between the given lyrics and corresponding melodies could be \textquotedblleft one-to-one \textquotedblright or \textquotedblleft one-to-many \textquotedblright. Most of previous methods only consider the \textquotedblleft one-to-one \textquotedblright alignment, which introduces bias into the melody composition.

To address the aforementioned issues, we propose a controllable song-level lyric-to-melody generation method called CSL-L2M, which is capable of generating melodies aligning with lyrics and user-specified musical attributes at the full-song level.
Specifically, we first introduce a novel music representation called REMI-Aligned. This representation incorporates strict syllable- and sentence-level lyric-melody alignments, which makes both exact and \textquotedblleft one-to-many \textquotedblright alignment learning feasible.
Inspired by \cite{wu2023musemorphose}, which equips conditional Transformer with the capability to model long sequences under fine-grained time-varying conditions through in-attention, we integrate the in-attention technique into our CSL-L2M model. Multiple multi-granularity lyric controls (including sentence-level semantic embeddings, word-level part-of-speech (POS) embeddings, and syllable-level tone embeddings) and musical controls (including coarse-level human-labeled musical tags, sentence-level statistical musical attributes, and learned high-fidelity musical features \cite{von2023figaro} from a pre-trained Vector Quantized-Variational AutoEncoder (VQ-VAE)) are extracted and fed into the conditional Transformer decoder through in-attention to realize tight fine-grained control of lyrics and musical attributes over melody generation process. This enables the generation of high-quality melodies from lyrics, precisely tailored to the user's desired musical style. Moreover, the musical controls not only enable user-controllable generation but also provide the model with additional musical information that is beneficial for melody modeling.
Experiments conducted on our 10,170 Chinese pop songs demonstrate that CSL-L2M could generate melodies that are both well-matched with the lyrics and consistent with the user-specified musical attributes. Compared to the state-of-the-art lyric-to-melody generation methods, CSL-L2M generates melodies with higher quality, better controllability and enhanced structure.

\section{Related Work}
\subsubsection{Lyric-to-Melody Generation}
The development of lyric-to-melody generation has evolved from traditional rule-based \cite{nichols2009lyric, monteith2012automatic} and statistical methods \cite{long2013t} to deep learning methods. The traditional methods usually rely on specific hand-designed musical rules and suffer from low generation quality. 
Currently, the end-to-end deep generative models are the mainstream methods but they suffer from several challenges: 
1) weak correlations between lyrics and melodies are difficult to capture by the network models, where much paired training data with alignment information is required;
2) strict alignment between each syllable in the given lyric and note in the corresponding melody is required, which needs additional alignment modeling. 
As for the first challenge, limited available paired lyric-melody data affects generation quality. The end-to-end models which directly learn the mapping from lyrics to melodies with the limited paired data often lead to poor generation quality. To this end,
SongMASS \cite{sheng2021songmass} improves the generation performance of end-to-end models by leveraging self-supervised pre-training on much unpaired lyric and melody data. Furthermore, TeleMelody \cite{ju2021telemelody}, a two-stage generation pipeline based on musical templates, is proposed to enhance data efficiency and further improve generation performance.
In addition, ReLyMe \cite{zhang2022relyme} introduces several principles of lyric-melody relationships from music theory into the decoding process, enhancing the harmony between lyrics and melodies. However, these methods fail to exploit melody-related lyric information and additional musical information for tightly controlling over the melody generation. Consequently, they are unable to adequately capture the intricate relationships between lyrics and melodies, resulting in limited generation quality. Moreover, few of them generate melodies at the full-song level, causing poor musical structure.
As for the second challenge, most existing works either focus solely on the \textquotedblleft one-to-one \textquotedblright lyric-melody alignment or do not ensure precise alignment, which can easily degrade generation quality.

\subsubsection{Controllable Music Generation}
Controllability in music generation aims to provide user control over the process in a desired direction \cite{briot2020deep}.
According to the levels of controllability, it can be divided into global/coarse-grained control and fine-grained control. The former refers to the fact that generation process is guided by time-invariant controls. Instead, the later refers to the fact that the generation process is guided by time-varying controls, which can provide more flexible and precise control, especially in the generation of long sequences.
Controllable music generation has attracted increasing research interest. In \cite{dong2018musegan, yang2017midinet, neves2022generating}, global conditions are injected into the training procedure of generative adversarial networks.
Some works \cite{payne2019musenet, sarmento2023gtr} achieve global control through conditional Transformer models with prompt-based control tokens. Many methods based on VAE enable users to exert global control by manipulating latent conditioning vectors \cite{brunner2018midi, roberts2018hierarchical,tan20music}. Transformer autoencoders are used in \cite{choi2020encoding} to realize improved control by learning global performance representations. 
However, these global controls often become less effective during long sequence generation, as the model may forget or weaken the global conditions over time.
In contrast, MuseMorphose \cite{wu2023musemorphose} and FIGARO \cite{von2023figaro} introduce fine-grained control, where the former is realized through one Transformer VAE based on an in-attention conditioning technique and the later is achieved through description-to-sequence learning.
Existing research on lyric-to-melody generation rarely pays attention to controllability. Only a few works have explored this area and do not offer fine-grained and flexible control.
In this paper, we delve into the controllability of lyric-to-melody generation.

\section{Methodology}
To overcome the difficulty of learning strict yet weak correlations between lyrics and melodies and enable user controls over full-song melody generation, we propose a controllable song-level lyric-to-melody generation method called CSL-L2M, as shown in Figure~\ref{fig_oview}. This method is capable of generating full-song melodies that match the given lyrics and adhere to user-specified musical attributes. 
We achieve this by employing the in-attention technique, as proposed in \cite{wu2023musemorphose}, to tightly control the conditional autoregressive Transformer decoder's generation process under multiple multi-granularity lyric and musical conditions.

\subsection{Technical Background}
The unconditional Transformer decoder's autoregressive generation process can be formulated as $p(x_t|x_{<t})$, where $x_t$ is the element of a sequence to predict at timestep $t$, and $x_{<t}$ represents all previously generated elements of the sequence. If a global condition vector $\bm{c}$ is offered to the model, the modeling could be formulated as $p(x_t|x_{<t},\bm{c})$. However, the global control tends to lose its effectiveness during long sequence generation. It is needed to incorporate fine-grained control mechanisms. Assuming that the target sequence consists of $N$ segments and each timestep index $t \in [1,T]$ belongs to one of the $N$ sets of indices $I_1, I_2,...,I_N$, where $I_n \cap I_{n^{\prime}}=\varnothing$ for $n \neq n^{\prime}$ and $\bigcup_{n=1}^{N} I_n=[1, T]$, the fine-grained control is achieved by providing the generation model with each segment-level condition vector $\bm{c}_n$ during the corresponding time interval $I_n$, formulated as:
\begin{equation}
p(x_t|x_{<t};\bm{c}_n), \quad \text{for} \; t\in I_n,
\label{eq:1}
\end{equation}
where the time-varying condition vectors $\bm{c}_1,\bm{c}_2,...,\bm{c}_N$ provide a high-level blueprint of the sequence to model.
This could be helpful for long sequence generation, particularly for full-song music generation.

There are many ways to condition autoregressive Transformer decoders at fine-grained level, where the in-attention conditioning \cite{wu2023musemorphose} offers tight control. Specifically, the in-attention method projects the segment-level condition vector $\bm{c}_n$ to the same space as the self-attention hidden stats via
\begin{equation}
	\bm{e}_n^{\top}=\bm{c}_n^{\top}{W}_{in}, \quad {W}_{in} \in \mathbb{R}^{d_c\times d}.
	\label{eq:2}
\end{equation}
Then the hidden condition state $\bm{e}_n$ is added to each hidden state of all the self-attention layers to obtain the input to the subsequent layer, formulated as:
\begin{equation}
\begin{split}
	\tilde{\bm{h}}_t^l &=\bm{h}_t^l+\bm{e}_n, \quad \forall l \in \{0,...,L-1\}; \\ 
	\bm{h}_t^{l+1} &=\text{SelfAttention}(\tilde{\bm{h}}_t^l),
\end{split}
\label{eq:3}
\end{equation}
which serves as a frequent reminder of the segment-level conditions for the Transformer decoder, thereby achieving tight control over the generation process.

\begin{figure}[t!]
	\centering
	\includegraphics[width=0.45\textwidth]{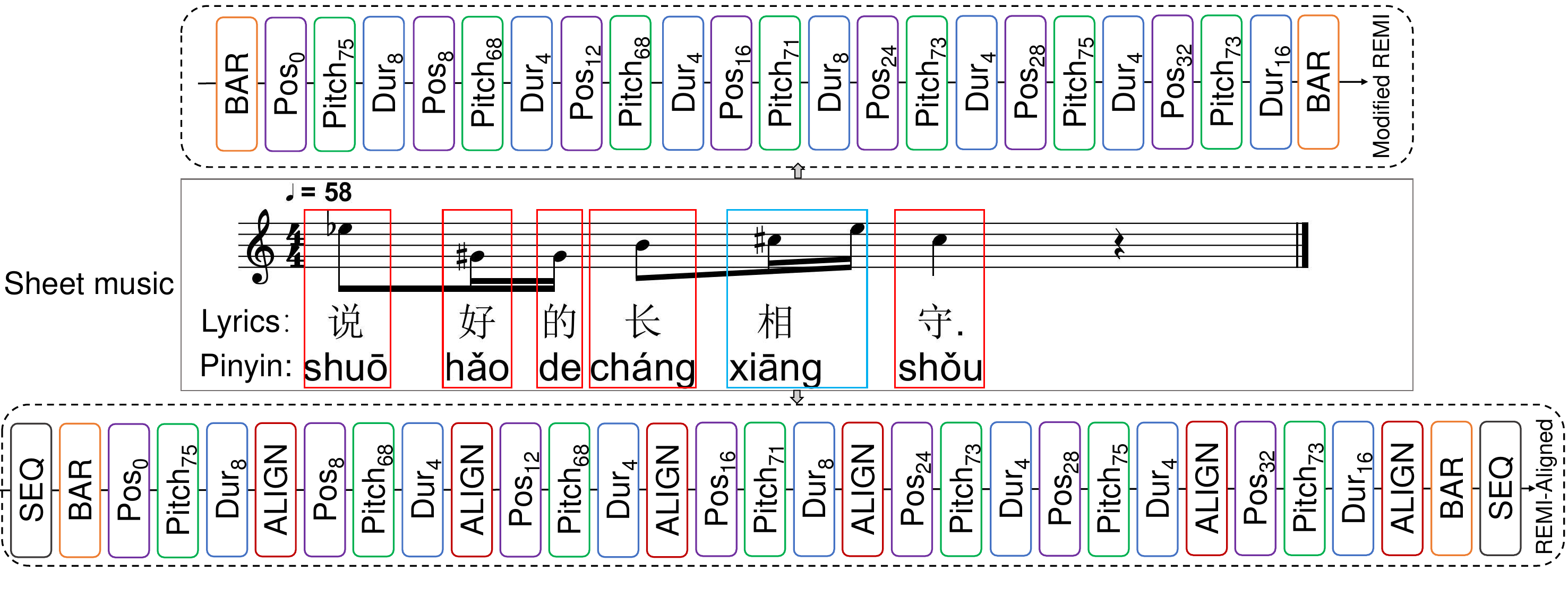} 
	\caption{Two representations of the same music piece.
	}
	\label{fig1}
\end{figure}

\begin{figure*}[t!]
	\centering
	\includegraphics[width=0.8\textwidth]{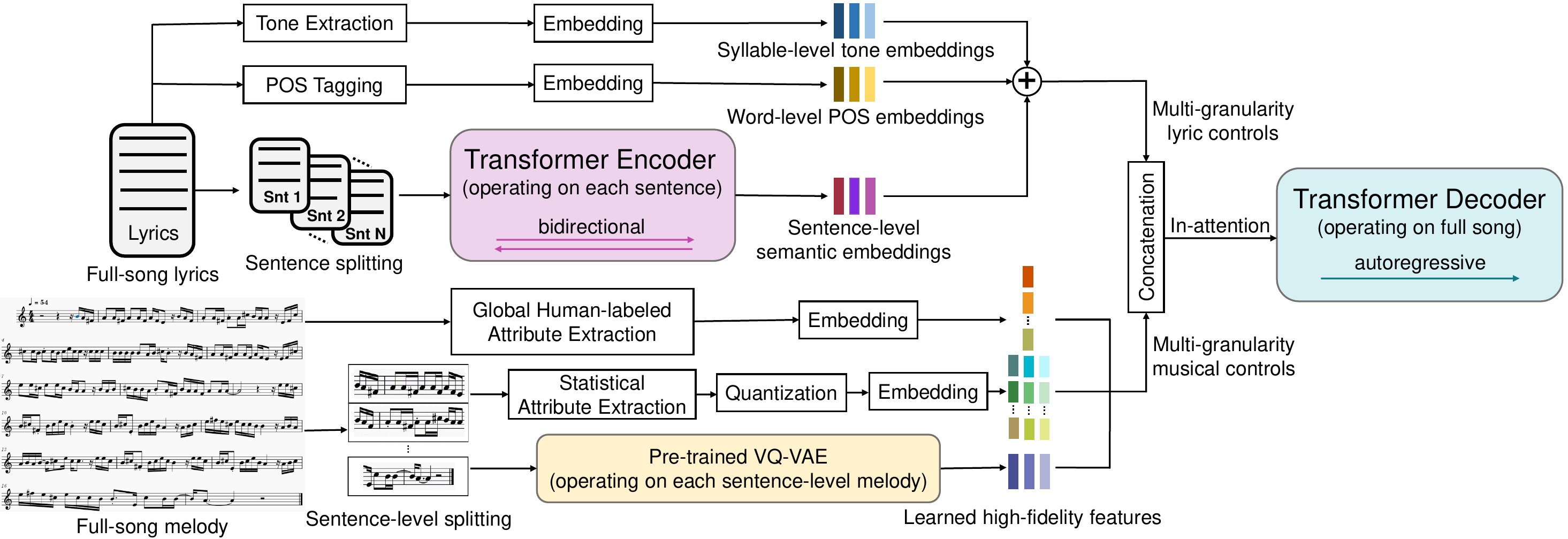} 
	\caption{Architecture of CSL-L2M.}
	\label{fig_oview}
\end{figure*}

\subsection{REMI-Aligned Representation}
To apply neural sequence models to symbolic music generation tasks, it is necessary to first convert a musical piece into a time-ordered sequence of discrete tokens. There are several ways to implement the conversion, leading to different sequence representations of the same music piece. One prevalent music representation is based on revamped MIDI-derived events (REMI) \cite{huang2020pop}. In REMI, a musical piece is represented as a time-ordered sequence of event tokens including bar, position, pitch, duration, velocity, tempo and chord. 

To precisely model the strict alignments between lyrics and melodies, we propose incorporating both sentence-level and syllable-level alignments into the music representation to enable explicit learning. Accordingly, we extend REMI to form a REMI-Aligned music representation by adding these alignments, while discarding tempo, chord and note velocity tokens given the fixed tempo in our dataset, irrelevance of note velocity to our task, and potential issues with chord accuracy. Furthermore, since 64th notes are the shortest notes in our dataset, we improve the temporal resolution of note position from 4 to 16 sub-beats per quarter note, enabling precise quantization of each note in our fixed 4/4 time signature dataset.
Examples of a music sequence encoded in modified REMI and REMI-Aligned are shown in Figure~\ref{fig1}.

\subsection{Model Architecture}
Figure~\ref{fig_oview} illustrates the architecture of our proposed CSL-L2M, consisting of a sentence-wise bidirectional Transformer encoder and an autoregressive Transformer decoder equipped with the capability to model full-song melodies under both lyric and musical multi-granularity controls. 

\subsubsection{Lyric Controls}
In CSL-L2M, we fully utilize the lyric information related to melodies in training, which improves the model capability to capture the correlations between lyrics and melodies. Specifically, we extract syllable-level tone embeddings, word-level POS embeddings, and sentence-level semantic embeddings serving as time-varying conditions at different granularities to exert fine-grained controls over the Transformer decoder's generation via the in-attention technique.

\textit{1) Tone:}
Tone \footnote{\url{https://en.wikipedia.org/wiki/Tone}}, in tonal languages, refers to the pitch variations that help distinguish words with the same spelling but different meanings, playing a crucial role in minimizing semantic ambiguities.
Around 60\% of languages have tone. In contrast to English, Chinese is a tonal language containing four main tones and one light tone in its characters. The pitch flow of a melody in Chinese songs is usually closely related to the tones of the corresponding lyrics. Accordingly, we incorporate the tone information into our model training to help learn pitch flow of the generated melodies to match with the given lyrics. The $s^{th}$ syllable-level tone attribute ${c}^{\text{tone}}_{s}$ of each song is converted to an embedding vector $\bm{c}^{\text{tone}}_{s}=\textbf{Emb}^{\text{tone}}({c}^{\text{tone}}_{s})$, before being fed into the decoder as a syllable-varying condition. 

\textit{2) POS:}
POS contains potential information of prosodic boundaries between words, which is helpful for enhancing rhythms and structures of generated melodies.
Consequently, the POS information is utilized for our model training. Specifically, we first perform POS tagging on the given lyrics by Jieba \footnote{\url{https://github.com/fxsjy/jieba}}, an open-source tool that supports 56 tags commonly used in Chinese. Then the $p^{th}$ word-level POS attribute ${c}^{\text{POS}}_{p}$ of each song is transformed into an embedding vector $\bm{c}^{\text{POS}}_{p}=\textbf{Emb}^{\text{POS}}({c}^{\text{POS}}_{p})$, before being fed into the decoder as a word-varying condition.

\textit{3) Semantic Embeddings:}
Since most previous works divide the input lyrics into sentences and then compose each piece of melody from the sentences one by one, we set the granularity of lyric text conditions to a sentence. We employ a bidirectional Transformer encoder \cite{vaswani2017attention} to learn to extract sentence-level latent semantic embeddings of the given lyrics, which is jointly trained with the Transformer decoder using the negative log-likelihood (NLL) training objective. More concretely, the input lyrics of each song are divided into sentences, formulated as $I=\{I_1,I_2,...,I_N\}$, where $I_n$ is the $n^{th}$ sentence of the lyrics. Then the Transformer encoder encodes these sentences in parallel. We treat the encoder's attention output at the first timestep (corresponding to the SEQ token of the sentence sequence tokens), i.e., $\bm{h}_{n,1}^{L_{\textbf{Enc}}}$, as the contextualized representation of the sequence. Finally, it is performed an affine transformation via a learnable weight $W$ to obtain the semantic embedding. These operations can be summarized as follows:
\begin{equation}
\begin{split}
\bm{h}_{n,1}^{L_{\textbf{Enc}}} &=\textbf{Enc}(I_n) \quad \text{for} \: 1\leq n \leq N; \\
\bm{z}_n^{\text{sem}} &={\bm{h}_{n,1}^{L_{\textbf{Enc}}}}^{\top}W, \quad {W} \in \mathbb{R}^{d\times d_l},
\end{split}
\label{eq:4}
\end{equation}
where $\bm{z}_n^{\text{sem}}$ is the semantic embedding for the $n^{th}$ sentence.

\subsubsection{Musical Controls}
To enable user control over the melody generation, we introduce human-labeled musical tags, sentence-level statistical musical attributes, and learned latent musical representations extracted from a pre-trained VQ-VAE serving as coarse-grained, fine-grained and high-fidelity controls, respectively, to the generation process.

\textit{1) Human-Labeled Musical Tags:}
We offer a high-quality, precisely annotated parallel lyric-melody dataset with alignment information consisting of 10,170 Chinese pop songs with time signature 4/4. Moreover, tags of key \footnote{\url{https://en.wikipedia.org/wiki/Key_(music)}}, emotion, and structure \footnote{\url{https://en.wikipedia.org/wiki/Song_structure}} of each song are meticulously annotated, encompassing 24 \footnote{12 major keys: C, D$\flat$, D, E$\flat$, E, F, F$\sharp$, G, A$\flat$, A, B$\flat$, B; \\ 12 minor keys: c, c$\sharp$, d, d$\sharp$, e, f, f$\sharp$, g, g$\sharp$, a, b$\flat$, b.}, 3 \footnote{3 emotions: Neutral, Positive, Negative.}, and 5 \footnote{5 structure sections: Verse, Chorus, Insertion, Bridge, Outro.} distinct categories respectively. 
The three types of musical tags serving as coarse-grained conditions are introduced into the decoder to realize human-interpretable control over the melody generation. Specifically, key is highly correlated with pitch distribution of the entire melody. The key of each song $c^{\text{key}}$ is transformed into an embedding vector by an embedding layer, i.e., $\bm{c}^{\text{key}}=\textbf{Emb}^{\text{key}}(c^{\text{key}})$ and then fed into the decoder as a global condition.
Similarly, the emotion of each song $c^{\text{emot}}$ is transformed into an embedding vector $\bm{c}^{\text{emot}}=\textbf{Emb}^{\text{emot}}(c^{\text{emot}})$ and then fed into the decoder as a global condition.
The verse-chorus form, serving as the cornerstone of pop songs, comprises two core sections---a verse and a chorus---that typically contrast melodically, rhythmically, harmonically and dynamically. We convert the $u^{th}$ structure-level attribute of each song into an embedding vector $\bm{c}^{\text{struc}}_u=\textbf{Emb}^{\text{struc}}(c^{\text{struc}}_u)$, and then fed it into the decoder as a structure-varying condition.

\textit{2) Statistical Musical Attributes:}
To enable fine-grained user controls over melody generation and help the model better learn correlations between lyrics and melodies, we introduce sentence-level statistical musical attributes. Their granularity is set to a sentence instead of a bar to maintain consistency with the semantic lyric embedding control granularity.
We utilize 12 types of statistical musical attributes: pitch mean (PM), pitch variance (PV), pitch range (PR), direction of melodic motion (DMM), amount of arpeggiation (AA), chromatic motion (CM), duration mean (DM), duration variance (DV), duration range (DR), prevalence of most common note duration (MCD), note density (ND), fraction of syllables in lyrics to notes in the corresponding melodies (Align) \footnote{Details of these musical attributes are available at \url{https://lichaiustc.github.io/CSL-L2M/}}. 
They are calculated for each sentence-level melody sequence. We first quantize these attributes into $K$ classes with approximately equal sample sizes, where $K=64$ in our work. Then the $n^{th}$ sentence-level attributes of the 12 statistical musical attributes for each song are converted into embedding vectors $\bm{c}_n^{\text{PM}}$, $\bm{c}_n^{\text{PV}}$, $\bm{c}_n^{\text{PR}}$, $\bm{c}_n^{\text{DMM}}$, $\bm{c}_n^{\text{AA}}$, $\bm{c}_n^{\text{CM}}$, $\bm{c}_n^{\text{DM}}$, $\bm{c}_n^{\text{DV}}$, $\bm{c}_n^{\text{DR}}$, $\bm{c}_n^{\text{MCD}}$, $\bm{c}_n^{\text{ND}}$, $\bm{c}_n^{\text{Align}}$, respectively and fed into the decoder. These controls can be grouped into four categories, i.e., pitch-related controls (pitch Ctls=$\text{Concat}(\bm{c}^{\text{PM}};\bm{c}^{\text{PV}};\bm{c}^{\text{PR}};\bm{c}^{\text{DMM}};\bm{c}^{\text{AA}};\bm{c}^{\text{CM}})$), duration-related controls (Dur Ctls=$\text{Concat}(\bm{c}^{\text{DM}};\bm{c}^{\text{DV}};\bm{c}^{\text{DR}};\bm{c}^{\text{MCD}})$), rhythm-related controls ($\bm{c}^{\text{ND}}$), and note-number-related controls ($\bm{c}^{\text{Align}}$).


\textit{3) Learned Musical Features:}
Inspired by \cite{von2023figaro}, we introduce learned musical features extracted from the latent space of a pre-trained VQ-VAE model to provide high-fidelity information to the decoder. This helps to alleviate non-injectivity problem in the lyric-to-melody generation task. The VQ-VAE model consists of a Transformer encoder, a Transformer decoder, and a vector quantization. For training, first, the full-song melody of each song is split into sentence-level melody sequences $X=\{X_1,X_2,...,X_N\}$, where $X_n$ is the $n^{th}$ sentence-level melody sequence and tokenized by REMI-Aligned. Then the Transformer encoder maps these sequences to the latent space in parallel. The encoder's attention output at the first timestep (corresponding to the SEQ token of the sentence-level melody sequence tokens) is considered as contextualized representation of the sequence. Finally, the quantized latent representations are obtained via the vector quantization and then fed into the decoder through in-attention to reconstruct the original full-song melody. Note that the hyperparameters in the vector quantization, i.e., latent group and codebook sizes, are set to 64 and 2048 respectively in our work.
Thus the sentence-level quantized latent representations $\bm{z}_n^{\text{learned}}$ extracted from the pre-trained VQ-VAE serving as learned musical features are introduce into our CSL-L2M training to provide high-fidelity information.

\subsubsection{Feeding Multi-Granularity Controls into the Transformer Decoder}
Borrowing the in-attention technique from \cite{wu2023musemorphose}, which conditions Transformer decoders with time-varying conditions during long sequence generation, we feed aforementioned multi-granularity controls into our Transformer decoder to achieve firm control over the full-song melody generation. Specifically, the word-level POS embeddings and sentence-level semantic embeddings are expanded to the syllable level by replication. Then they are added to the tone embeddings to get the syllable-level lyric-related controls $\bm{c}_s^{\text{lyric}}=\bm{c}_s^{\text{tone}}+\bm{c}_s^{\text{POS}}+\bm{z}_s^{\text{sem}}$. Next, aforementioned multi-granularity musical controls are expanded to the syllable level by replication according to the alignment information between lyrics and melodies. Finally, these syllable-level lyric and musical controls are fed into the decoder through in-attention after concatenation, i.e., 
\begin{equation}
\begin{split}
\bm{c}_s &=\text{concat}([\bm{c}_s^{\text{lyric}};\bm{c}_s^{\text{key}};\bm{c}_s^{\text{emot}};\bm{c}_s^{\text{struc}};\bm{c}_s^{\text{PM}};\bm{c}_s^{\text{PV}};\bm{c}_s^{\text{PR}}; \bm{c}_s^{\text{DMM}};\\
&\hphantom{tee}\bm{c}_s^{\text{AA}};\bm{c}_s^{\text{CM}};\bm{c}_s^{\text{DM}};\bm{c}_s^{\text{DV}};\bm{c}_s^{\text{DR}};\bm{c}_s^{\text{MCD}};\bm{c}_s^{\text{ND}};\bm{c}_s^{\text{Align}};\bm{z}_s^{\text{learned}}]), \\
\bm{y}_t&=\textbf{Dec}(x_{<t};\bm{c}_s).
\end{split}
\label{eq:5}
\end{equation}

\begin{figure}[t!]
	\centering
	\begin{minipage}{0.495\linewidth}
		\centering
		\includegraphics[width=0.99\linewidth]{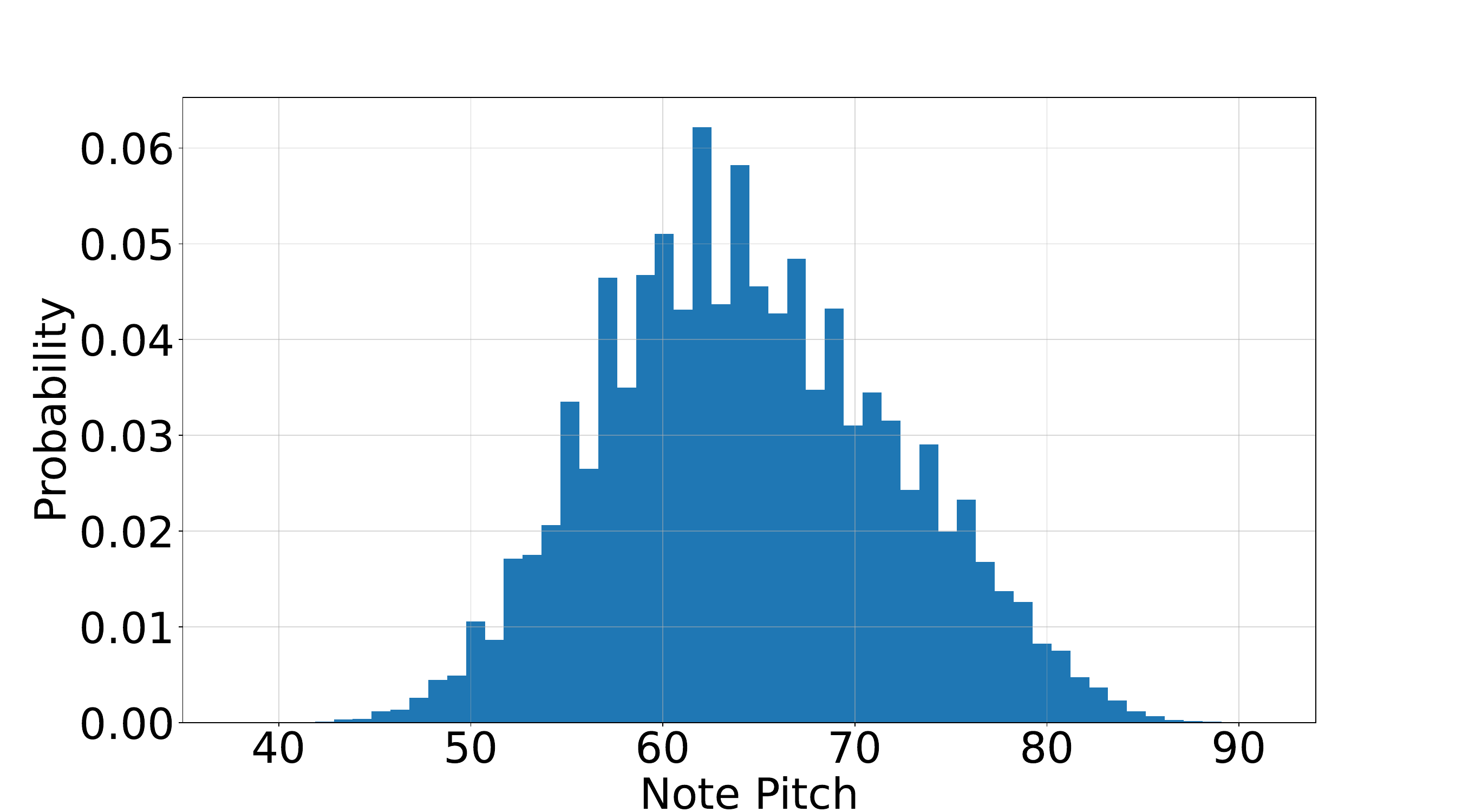}
	\end{minipage}
	\begin{minipage}{0.495\linewidth}
		\centering
		\includegraphics[width=0.99\linewidth]{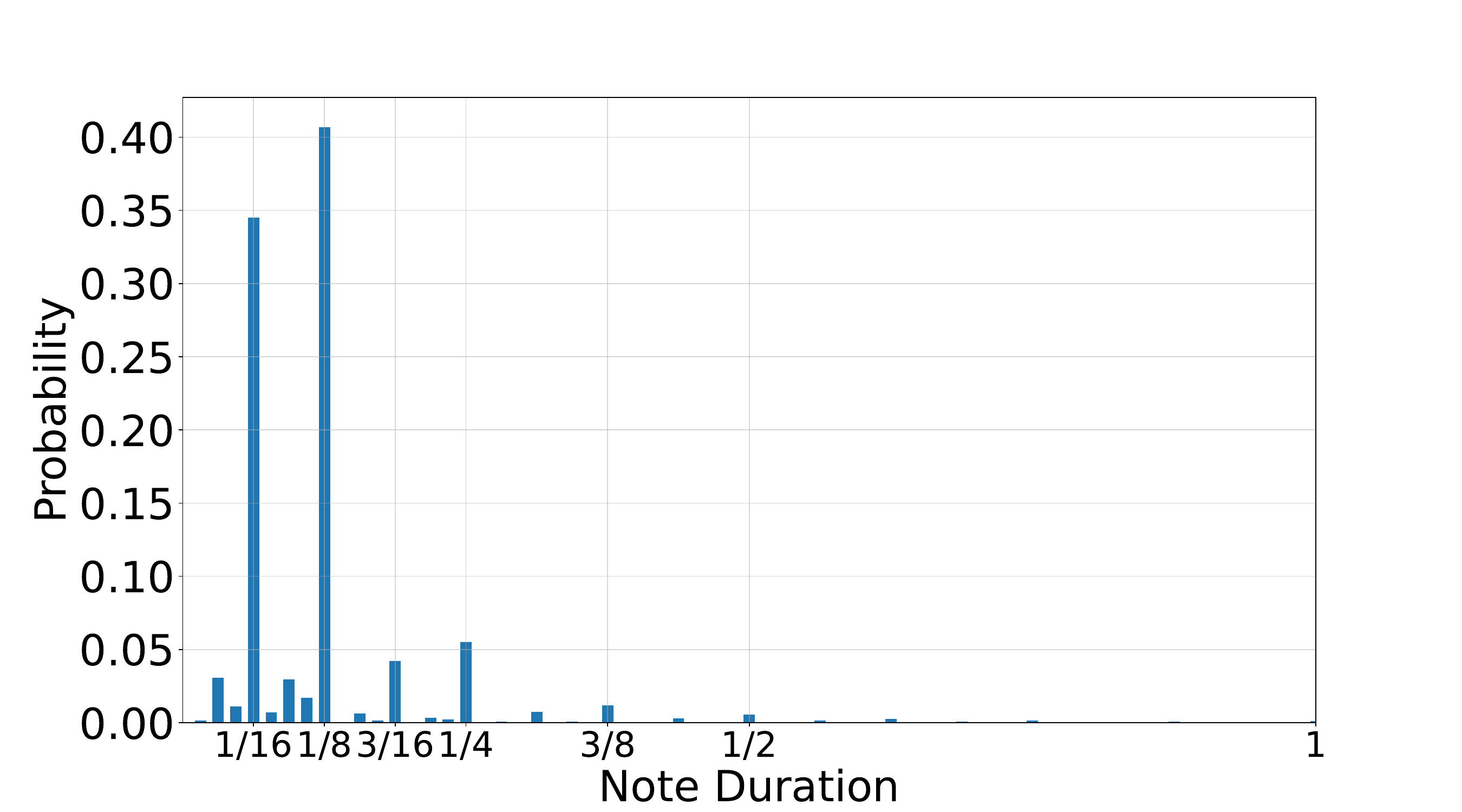}
	\end{minipage}
	\begin{minipage}{0.495\linewidth}
		\centering
		\includegraphics[width=0.99\linewidth]{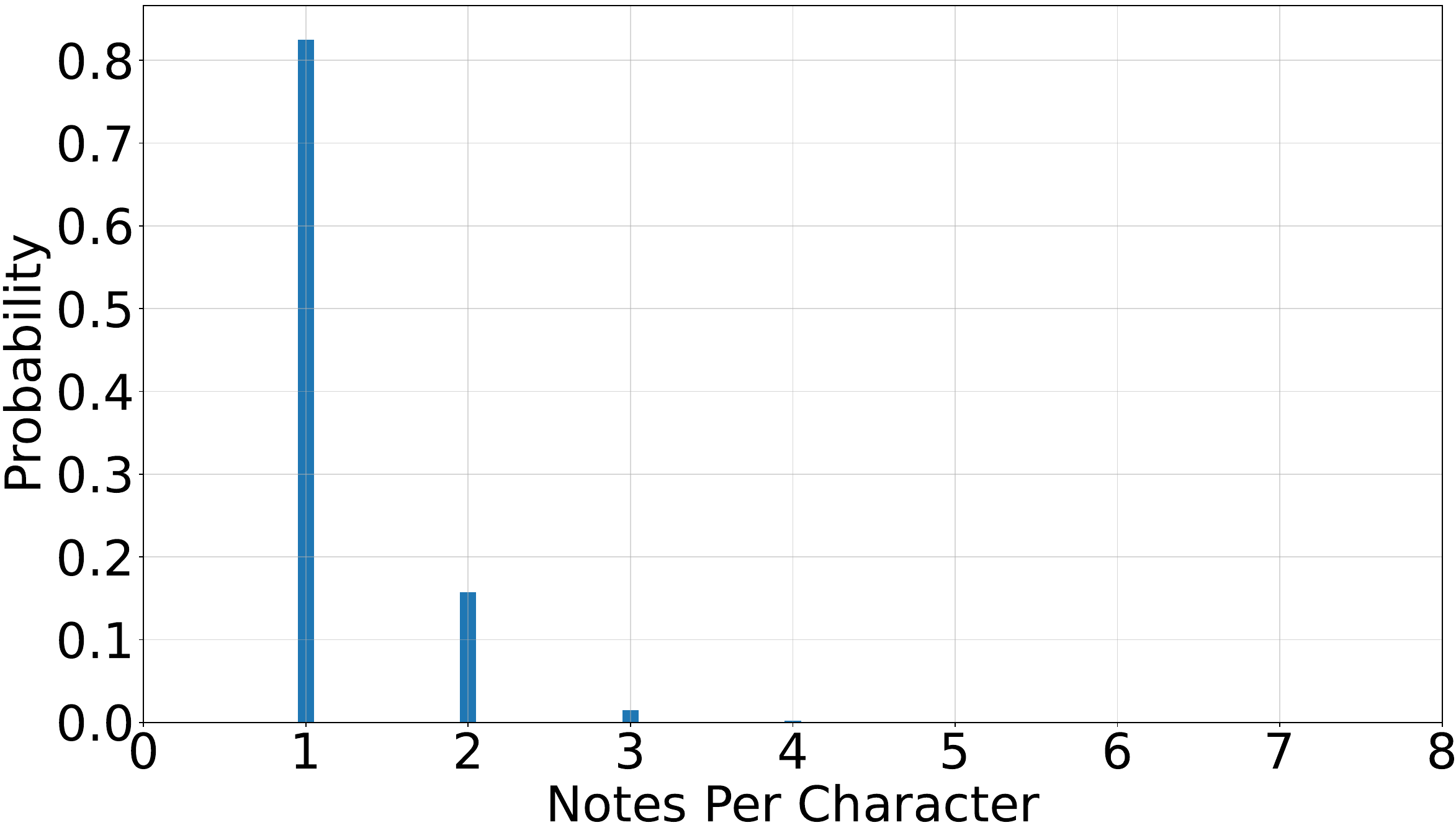}
	\end{minipage}
	\begin{minipage}{0.495\linewidth}
		\centering
		\includegraphics[width=0.99\linewidth]{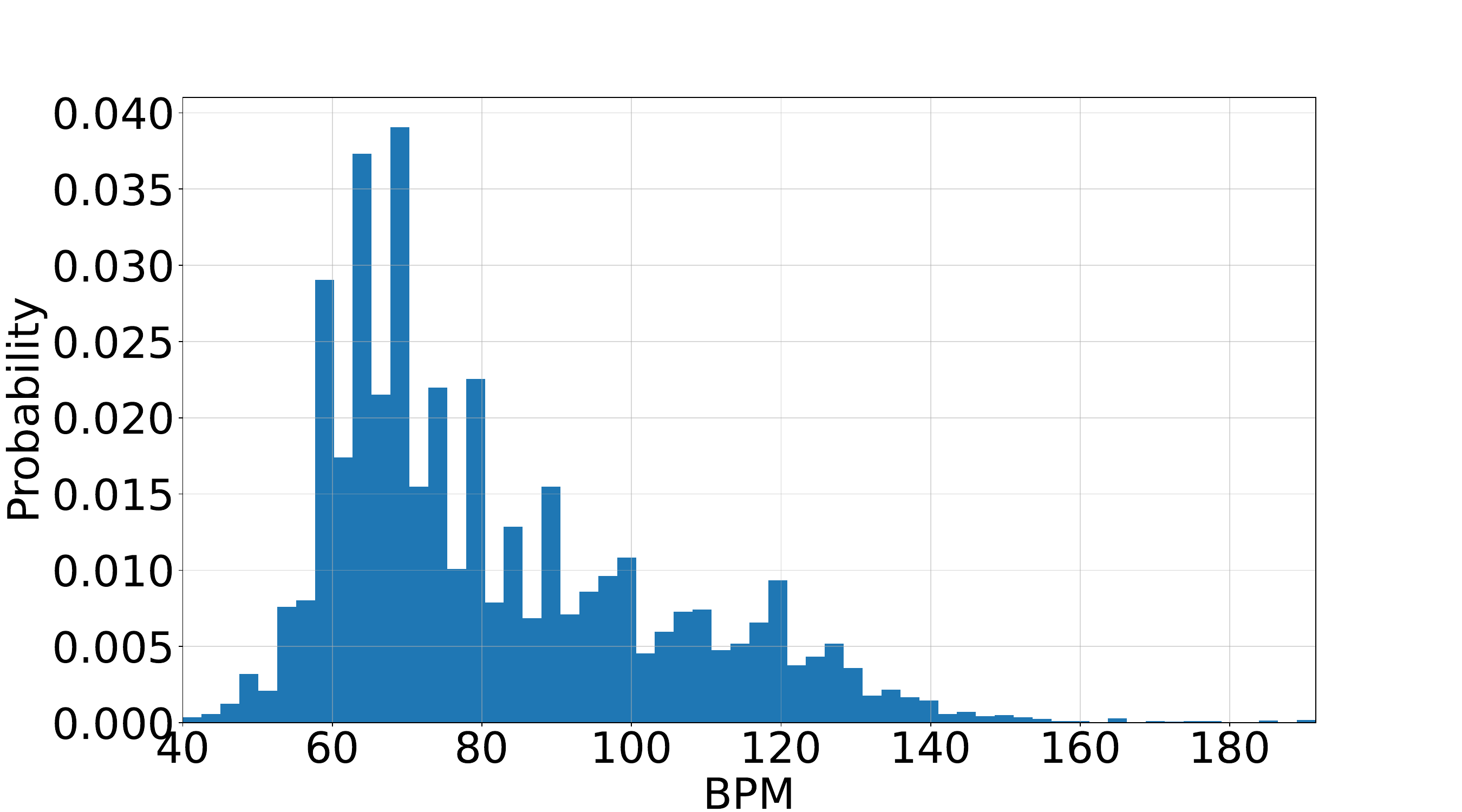}
	\end{minipage}
	\caption{Distributions of music attributes in our paired lyric-melody dataset. }
	\label{fig_data}
\end{figure}

\begin{table*}[t!]
	\centering
	\begin{adjustbox}{max width=0.78\textwidth}
		\begin{tabular}{l|ccc|ccccc}
			\toprule
			\multirow{2}{*}{Model} & \multicolumn{3}{c|}{Objective} & \multicolumn{5}{c}{Subjective} \\
			\cmidrule(lr){2-4}\cmidrule(l){5-9}
			& PD(\%)$\uparrow$ & DD(\%)$\uparrow$ & MD$\downarrow$ & Harmony$\uparrow$ & Rhythm$\uparrow$ & Structure$\uparrow$ & Emotion$\uparrow$ & Quality$\uparrow$ \\
			\midrule
			SongComposer \cite{ding2024songcomposer} & 33.37 & 44.98 & 3.12  & 2.32 & 2.56 & 2.33 & 2.51 & 2.47 \\
			TeleMelody \cite{ju2021telemelody} & 40.02 & 49.82 & 2.93  & 2.71  & 2.90 & 2.45 & 2.42 & 2.72 \\
			\midrule
			CSL-L2M & 86.35 & 93.50 & 1.27  & 3.74  & 4.03 & 4.20 & 3.86 & 3.94 \\
			\bottomrule
		\end{tabular}
	\end{adjustbox}
	\caption{Objective and subjective evaluation results of our CSL-L2M and compared models.}
	\label{tab:results}
\end{table*}

\section{Experiments}
\subsection{Experimental Settings}
\subsubsection{Dataset}
The collection of paired lyric-melody data is difficult due to the need for precise synchronization between lyrics and melodies as shown in the sheet music in Figure~\ref{fig1}, which requires detailed annotation and specific expertise. Currently, available paired lyric-melody dataset with alignment information is limited and of insufficient quality. To this end, we offer a high-quality, precisely annotated parallel lyric-melody dataset, encompassing 10,170 Chinese pop songs with time signature 4/4. 
Moreover, musical tags for each song including key, lyric emotion, song structure, and beats per minute (BPM) are precisely annotated.
We perform some statistics on this dataset shown in Figure~\ref{fig_data}. The following observations are made: 1) the most pitch/MIDI numbers fall within the range of 50 to 80; 2) in contrast to melodies in the English dataset \cite{yu2021conditional}, the melodies in our Chinese dataset feature a predominance of short musical notes, specifically 8th and 16th notes; 3) more than 80\% of characters/syllables correspond to a single musical note (i.e. \textquotedblleft one-to-one \textquotedblright alignment), and nearly 20\% of characters correspond to two or more notes (i.e.  \textquotedblleft one-to-many \textquotedblright alignment); 4) the BPM of most songs falls within the range of 60 to 120.
The 10,170 Chinese pop songs are split into the training, validation, and test sets in an 9:0.5:0.5 ratio for our experiments.

\subsubsection{Implementation Details}
Both the encoder and decoder of our CSL-L2M and VQ-VAE models consist of 12 self-attention layers with 8 self-attention heads, 512 hidden size and 2048 feed-forward dimension.
The dimension of each lyric attribute embedding as well as learned musical feature is 128. And the dimension of each human-annotated and statistical musical attribute embedding is 32. The models are trained with Adam optimizer and teacher forcing. We use linear warm-up to increase the learning rate to $10^{-4}$ in the first 200 steps, followed by a 150k-step cosine decay down to $5\times 10^{-6}$. The batch size is set to 4. During inference, nucleus sampling \cite{holtzman2019curious} is used to sample from the decoder output distribution at each timestep with a softmax temperature $\tau=1.2$ and truncating the distribution at cumulative probability $p=0.9$.

\subsubsection{Evaluation Metrics}
Unlike previous works that evaluate generated melodies from lyrics at the sentence level, we conduct evaluations on full-song melodies.

\textit{1) Objective Metrics:}
Objective evaluations are conducted on our test set comprising around 500 songs from our 10,170 Chinese pop songs. We focus on assessing the similarity between the generated and the ground-truth melodies. The following objective metrics proposed by \cite{sheng2021songmass} are adopted: 1) Pitch Distribution Similarity (PD); 2) Duration Distribution Similarity (DD); 3) Melody Distance (MD).

\textit{2) Subjective Metrics:}
Subjective evaluations are conducted on 10 songs randomly selected from our test set. We invite 70 participants (including 50 amateurs and 20 professionals) to score the melody properties using a scale from 1 (Poor) to 5 (Perfect). The following subjective metrics are considered: 1) Harmony: Is the melody itself harmonious as well as harmonized with the lyrics ? 2) Rhythm: Does the rhythm sound natural and match the rhythm of the lyrics? 3) Structure: How well does the melody structure match lyric structure? Specifically, whether lyrics with similar rhythm patterns have similar melodies? Does the melody feature a distinguishable verse-chorus structure? Are the transitions between contiguous phrases natural and coherent?
4) Emotion: Does the melody convey a consistent emotion with the lyrics? 5) Quality: What is the overall quality of the melody?

\begin{table}[t!]
	\centering
	\begin{adjustbox}{max width=0.4\textwidth}
		\begin{tabular}{l|ccc}
			\toprule
			& PD(\%)$\uparrow$ & DD(\%)$\uparrow$ & MD$\downarrow$\\
			\midrule
			CSL-L2M & 86.35 & 93.50 & 1.27  \\
			\midrule
			+ w/ learned Ctls & 97.98  & 98.62 & 0.25  \\
			- w/o Dur Ctls & 85.82 & 86.41 & 1.37   \\
			- w/o Dur+pitch Ctls & 66.07 & 85.65 & 1.78  \\
			- w/o Dur+pitch+ND+Align Ctls & 63.18 & 63.59 & 2.03   \\
			- w/o musical Ctls & 49.20 & 59.13 & 2.26  \\
			\bottomrule
		\end{tabular}
	\end{adjustbox}
	\caption{Objective evaluation results of our CSL-L2M under different controls.}
	\label{tab:results2}
\end{table}

\begin{figure}[t!]
	\centering
	\begin{minipage}{0.48\linewidth}
		\centering
		\includegraphics[width=0.99\linewidth]{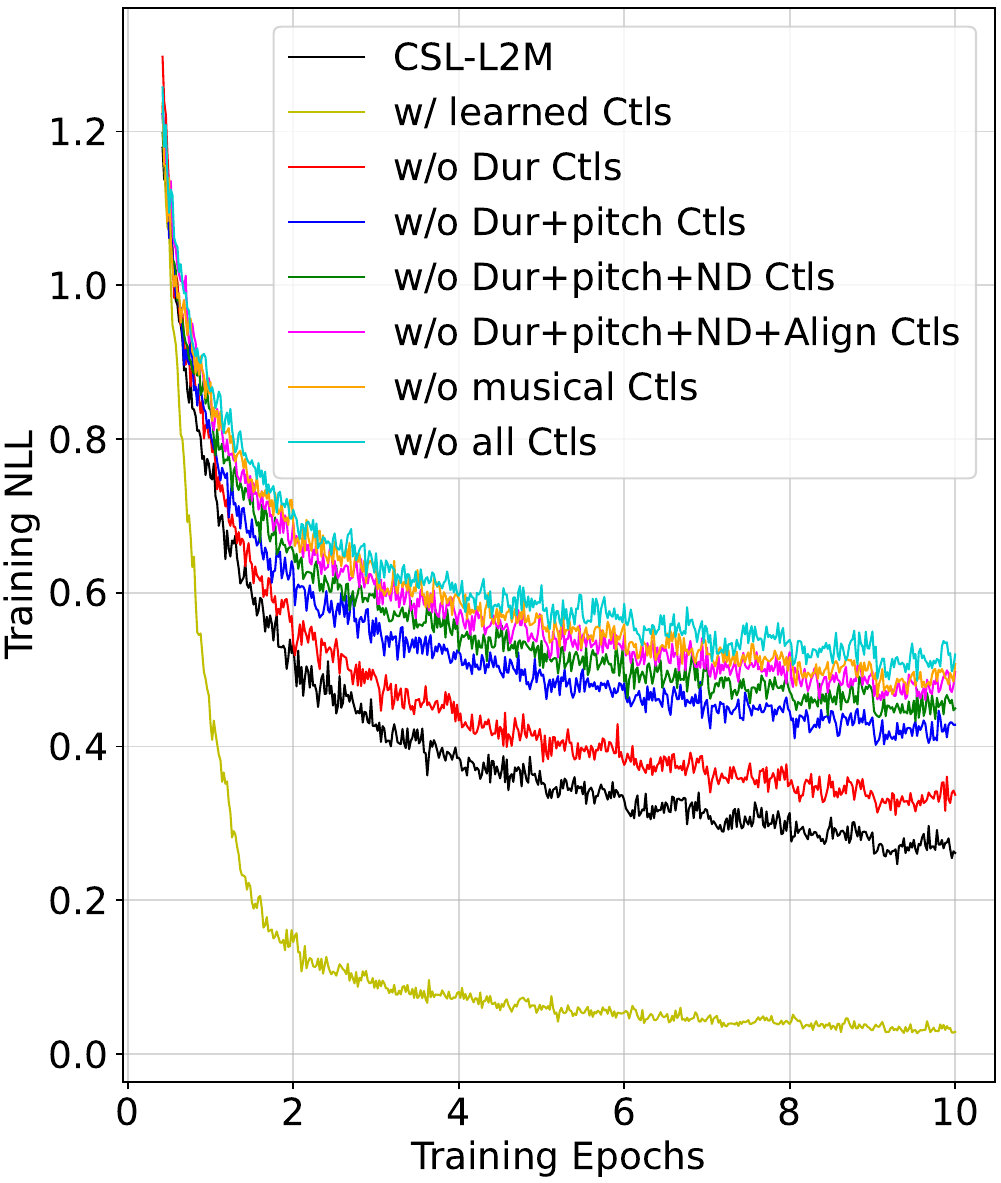}
	\end{minipage}
	\begin{minipage}{0.48\linewidth}
		\centering
		\includegraphics[width=0.99\linewidth]{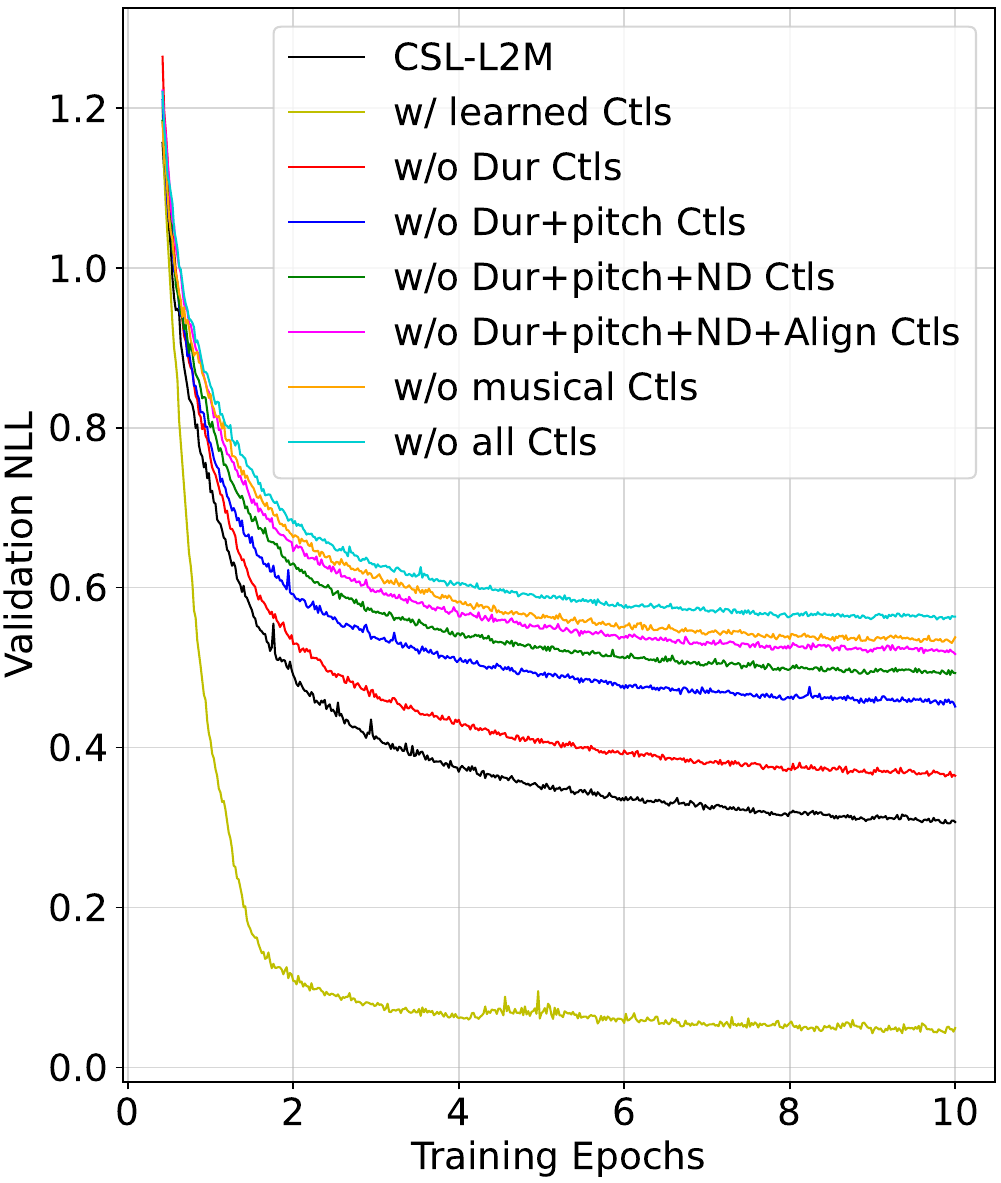}
	\end{minipage}
	\caption{Training dynamics of our CSL-L2M under different controls. }
	\label{fig_loss}
\end{figure}

\subsection{Experimental Results}
\subsubsection{Main Results}
Since learned musical features need to be extracted from existing melodies, unless otherwise specified, our reference to CSL-L2M refers to the version without learned musical controls. The evaluation of the full version will be conducted later in the context of style transfer and controllable generation.
We first compare our CSL-L2M with two state-of-the-art models, i.e. TeleMelody \cite{ju2021telemelody} and SongComposer \cite{ding2024songcomposer}. As shown in Table~\ref{tab:results}, CSL-L2M significantly outperforms advanced models, namely SongComposer and TeleMelody, in both objective and subjective evaluations, demonstrating the effectiveness of CSL-L2M in generating high-quality song-level melodies from lyrics. We further perform ablation study to verify the effectiveness of lyric and musical controls in CSL-L2M. As illustrated in Table~\ref{tab:results2} and Figure~\ref{fig_loss}, by successively removing duration-related controls, pitch-related controls, note density and alignment controls, and human-annotated musical attribute controls, we observe a continuous performance degradation. This indicates that the musical controls, in addition to enabling user-controlled generation, can help conditional Transformer in modeling melodies because they offer the model more musical information for reference.
Besides, the learned musical features include high-fidelity information of melodies, which aids in reducing non-injectivity of the generation model. As a result, we find that CSL-L2M equipped with learned musical controls achieves top performance that nearly reaches the ceiling. Moreover, CSL-L2M with only lyric controls exceeds the performance of the two state-of-the-art models. This confirms the effectiveness of our designed fine-grained lyric controls and the lyric-to-melody generation framework based on conditional Transformer with the in-attention conditioning mechanism.

\subsubsection{Controllability Study}
\begin{figure}[t!]
	\centering
	\includegraphics[width=0.44\textwidth]{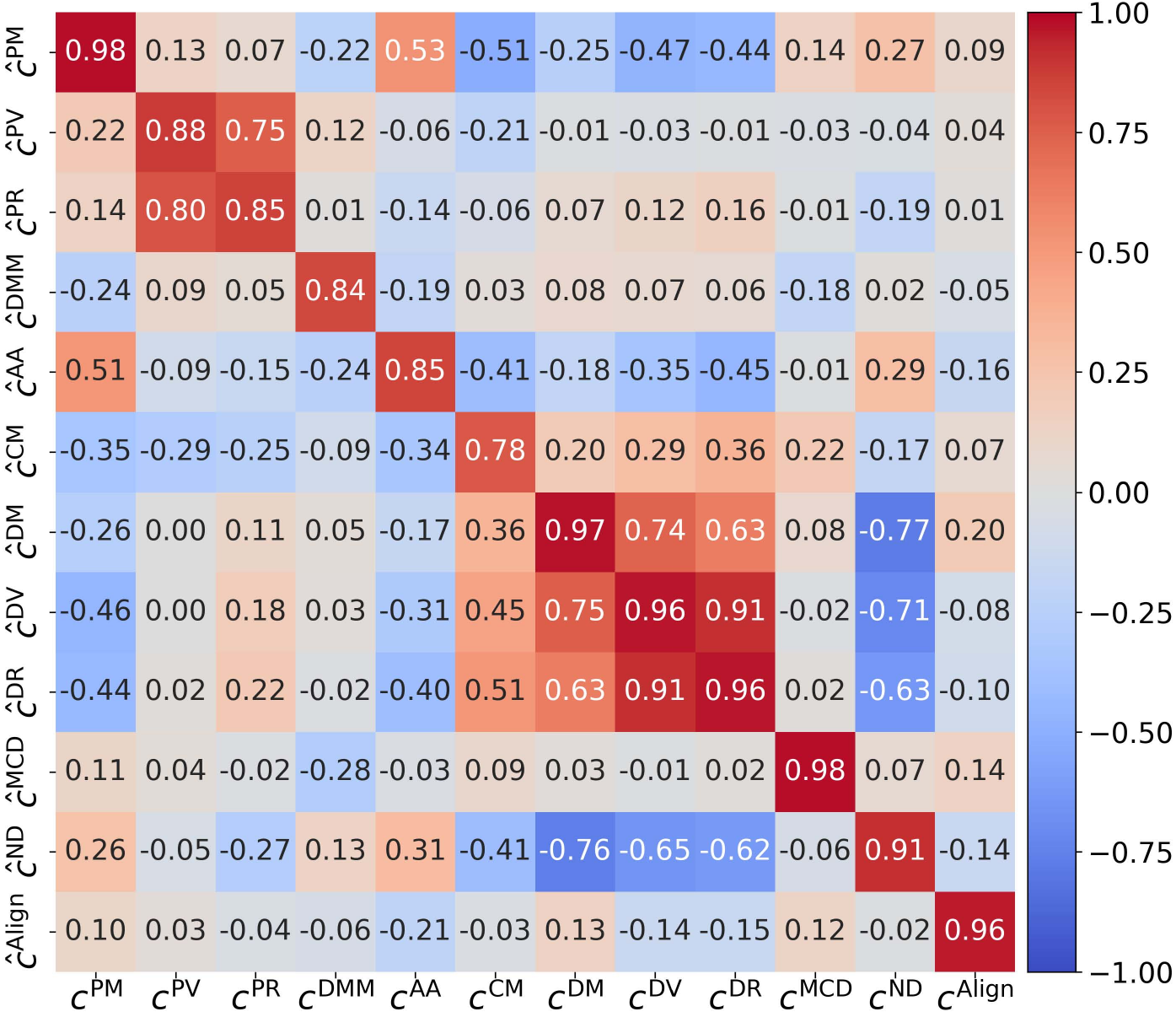} 
	\caption{Spearman’s rank correlation coefficients between user-specified attribute classes and the attribute raw classes computed from the generated melodies.}
	\label{fig_rho}
\end{figure}

Given that our statistical musical controls are ordinal by nature, following \cite{kawai2020attributes} and \cite{wu2023musemorphose}, we use the Spearman’s rank correlation coefficient $\rho$ to quantitatively assess the strength of statistical musical attribute control. To simultaneously evaluate the impact on other unrelated attributes when transferring a specific attribute, we calculate Spearman’s rank correlation coefficient matrix between the user-specified attribute classes and attribute raw scores derived from the generated melodies.
Results in Figure~\ref{fig_rho} reveal the strong and independent controllability strengths of CSL-L2M in attribute control. Specifically, for example, $\rho_{\text{PM}}=0.98$ denotes a strong and positive correlation between the user-specified attribute class $c^{\text{PM}}$ and the attribute raw class $\hat{c}^{\text{PM}}$ computed from the generated melodies, which demonstrates strong controllability of the pitch mean attribute. In contrast, $\rho_{\text{PM}|\text{Align}}=0.09$ is the correlation coefficient between the user-specified alignment attribute class $c^{\text{Align}}$ and the unrelated attribute class $\hat{c}^{\text{PM}}$ computed from the generated melodies, revealing the independent controllability of attributes in the multi-attribute scenario.

\subsubsection{Case Study}
\begin{figure}[t!]
	\centering
	\begin{subfigure}{0.47\textwidth}
		\includegraphics[width=\textwidth]{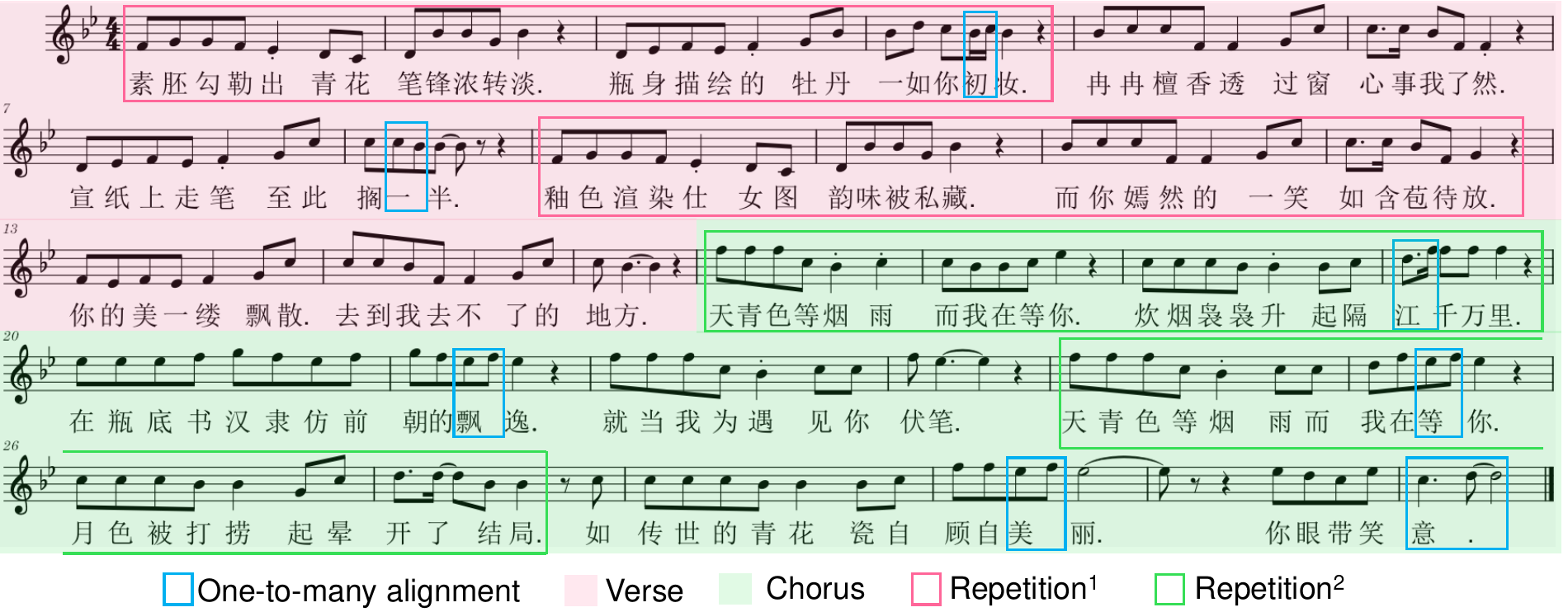}
		\caption{High-quality and well-structured generation.}
		\label{fig:sub1}
	\end{subfigure}
	\hfill
	\begin{subfigure}{0.47\textwidth}
		\includegraphics[width=\textwidth]{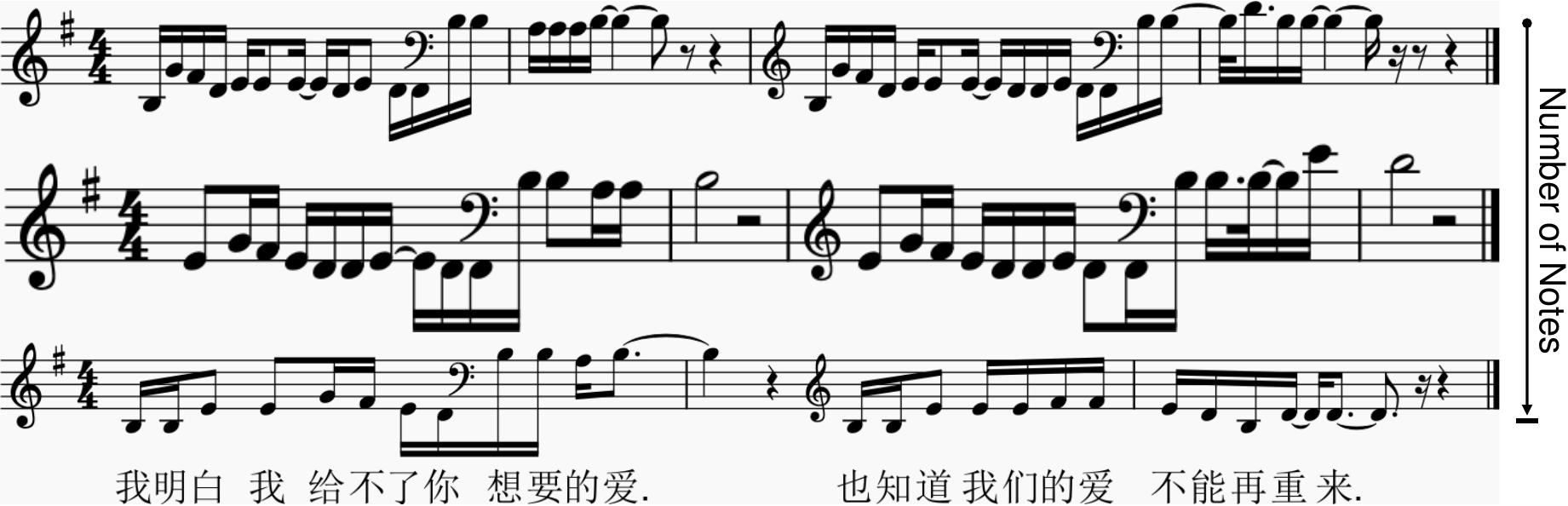}
		\caption{Strong controllability.}
		\label{fig:sub2}
	\end{subfigure}
	\hfill
	\begin{subfigure}{0.47\textwidth}
	\includegraphics[width=\textwidth]{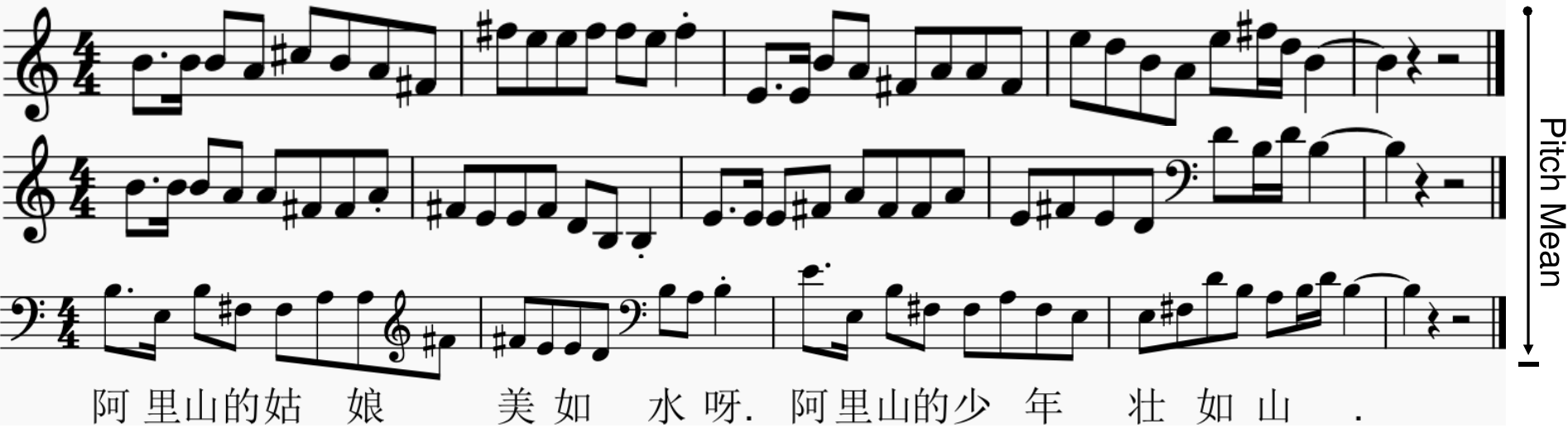}
	\caption{High-fidelity controllability.}
	\label{fig:sub3}
	\end{subfigure}
	\caption{Case study on generated melodies from CSL-L2M.}
	\label{fig_struct}
\end{figure}

In Figure~\ref{fig_struct}, we present some generated sheet music given lyrics to demonstrate the advantages of our CSL-L2M in terms of generation quality and controllability \footnote{More demos are available through \url{https://lichaiustc.github.io/CSL-L2M/}}. Due to space constraint, only a portion of the full-song melody is provided here.
Specifically, Figure~\ref{fig:sub1} shows that the generated melodies not only harmonize with given lyrics but also exhibit a coherent and distinguishable verse-chorus structure, along with repetition patterns matching lyrics. Besides, it is observed that our model can well model the "one-to-many" alignment relationship between lyrics and melodies. In Figure~\ref{fig:sub2}, the generate melodies well adhere to user-specified musical attributes. Figure~\ref{fig:sub3} presents high-fidelity style transfer results of CSL-L2M equipped with the learned musical features, confirming that the learned features provide high-fidelity melody information to the Transformer decoder. 
In summary, our proposed CSL-L2M could generate melodies that not only match with the given lyrics but also adhere to user-specified musical attributes.

\section{Conclusion}
To address weak controllability, low-quality and poorly structured generation issues in the lyric-to-melody generation task, we propose CSL-L2M in this paper towards controllable song-level melody generation conditioning on lyrics and user-specified musical attributes. We first introduce a novel music representation named REMI-Aligned to facilitate precise lyric-melody alignment relationship modeling. Then 
multiple multi-granularity lyric and musical attribute controls are extracted and fed into the conditional Transformer decoder through in-attention to achieve firm control over the generation process. Experiments demonstrate that our proposed CSL-L2M outperforms the state-of-the-art models in terms of generation quality and controllability. We believe our contributions will further advance the under-explored field of lyric-to-melody generation.

\section{Acknowledgments}
This work was supported by the National Science and Technology Innovation 2030 – Major Project (Grant No.2022ZD0208800) and the National Natural Science Foundation of China (Grant No. 62176215).

\bibliography{aaai25}

\begin{thebibliography}{32}
\providecommand{\natexlab}[1]{#1}

\bibitem[{Bao et~al.(2019)Bao, Huang, Wei, Cui, Wu, Tan, Piao, and
  Zhou}]{bao2019neural}
Bao, H.; Huang, S.; Wei, F.; Cui, L.; Wu, Y.; Tan, C.; Piao, S.; and Zhou, M.
  2019.
\newblock Neural melody composition from lyrics.
\newblock In \emph{Natural Language Processing and Chinese Computing: 8th CCF
  International Conference, NLPCC 2019, Dunhuang, China, October 9--14, 2019,
  Proceedings, Part I 8}, 499--511. Springer.

\bibitem[{Briot and Pachet(2020)}]{briot2020deep}
Briot, J.-P.; and Pachet, F. 2020.
\newblock Deep learning for music generation: challenges and directions.
\newblock \emph{Neural Computing and Applications}, 32(4): 981--993.

\bibitem[{Brunner et~al.(2018)Brunner, Konrad, Wang, and
  Wattenhofer}]{brunner2018midi}
Brunner, G.; Konrad, A.; Wang, Y.; and Wattenhofer, R. 2018.
\newblock MIDI-VAE: Modeling dynamics and instrumentation of music with
  applications to style transfer.
\newblock In \emph{ISMIR}, 747--754.

\bibitem[{Choi et~al.(2020)Choi, Hawthorne, Simon, Dinculescu, and
  Engel}]{choi2020encoding}
Choi, K.; Hawthorne, C.; Simon, I.; Dinculescu, M.; and Engel, J. 2020.
\newblock Encoding musical style with transformer autoencoders.
\newblock In \emph{International conference on machine learning}, 1899--1908.
  PMLR.

\bibitem[{Ding et~al.(2024)Ding, Liu, Dong, Zhang, Qian, He, Lin, and
  Wang}]{ding2024songcomposer}
Ding, S.; Liu, Z.; Dong, X.; Zhang, P.; Qian, R.; He, C.; Lin, D.; and Wang, J.
  2024.
\newblock Songcomposer: A large language model for lyric and melody composition
  in song generation.
\newblock \emph{arXiv preprint arXiv:2402.17645}.

\bibitem[{Dong et~al.(2018)Dong, Hsiao, Yang, and Yang}]{dong2018musegan}
Dong, H.-W.; Hsiao, W.-Y.; Yang, L.-C.; and Yang, Y.-H. 2018.
\newblock Musegan: Multi-track sequential generative adversarial networks for
  symbolic music generation and accompaniment.
\newblock In \emph{Proceedings of the AAAI Conference on Artificial
  Intelligence}, volume~32.

\bibitem[{Duan, Yu, and Oyama(2024)}]{duan2024semantic}
Duan, W.; Yu, Y.; and Oyama, K. 2024.
\newblock Semantic dependency network for lyrics generation from melody.
\newblock \emph{Neural Computing and Applications}, 36(8): 4059--4069.

\bibitem[{Duan et~al.(2022)Duan, Zhang, Yu, and Oyama}]{duan2022interpretable}
Duan, W.; Zhang, Z.; Yu, Y.; and Oyama, K. 2022.
\newblock Interpretable melody generation from lyrics with discrete-valued
  adversarial training.
\newblock In \emph{Proceedings of the 30th ACM international conference on
  multimedia}, 6973--6975.

\bibitem[{Hahn et~al.(2023)Hahn, Zhu, Mak, Rudin, and
  Jiang}]{hahn2023interpretable}
Hahn, S.; Zhu, R.; Mak, S.; Rudin, C.; and Jiang, Y. 2023.
\newblock An Interpretable, Flexible, and Interactive Probabilistic Framework
  for Melody Generation.
\newblock In \emph{Proceedings of the 29th ACM SIGKDD Conference on Knowledge
  Discovery and Data Mining}, 4089--4099.

\bibitem[{Holtzman et~al.(2020)Holtzman, Buys, Du, Forbes, and
  Choi}]{holtzman2019curious}
Holtzman, A.; Buys, J.; Du, L.; Forbes, M.; and Choi, Y. 2020.
\newblock The curious case of neural text degeneration.
\newblock \emph{The Eighth International Conference on Learning
  Representations}.

\bibitem[{Huang and Yang(2020)}]{huang2020pop}
Huang, Y.-S.; and Yang, Y.-H. 2020.
\newblock Pop music transformer: Beat-based modeling and generation of
  expressive pop piano compositions.
\newblock In \emph{Proceedings of the 28th ACM international conference on
  multimedia}, 1180--1188.

\bibitem[{Ju et~al.(2021)Ju, Lu, Tan, Wang, Zhang, Wu, Zhang, Li, Qin, and
  Liu}]{ju2021telemelody}
Ju, Z.; Lu, P.; Tan, X.; Wang, R.; Zhang, C.; Wu, S.; Zhang, K.; Li, X.; Qin,
  T.; and Liu, T.-Y. 2021.
\newblock Telemelody: Lyric-to-melody generation with a template-based
  two-stage method.
\newblock \emph{arXiv preprint arXiv:2109.09617}.

\bibitem[{Kawai, Esling, and Harada(2020)}]{kawai2020attributes}
Kawai, L.; Esling, P.; and Harada, T. 2020.
\newblock Attributes-Aware Deep Music Transformation.
\newblock In \emph{ISMIR}, 670--677.

\bibitem[{Long, Wong, and Sze(2013)}]{long2013t}
Long, C.; Wong, R. C.-W.; and Sze, R. K.~W. 2013.
\newblock T-music: A melody composer based on frequent pattern mining.
\newblock In \emph{2013 IEEE 29th International Conference on Data Engineering
  (ICDE)}, 1332--1335. IEEE.

\bibitem[{Monteith, Martinez, and Ventura(2012)}]{monteith2012automatic}
Monteith, K.; Martinez, T.~R.; and Ventura, D. 2012.
\newblock Automatic Generation of Melodic Accompaniments for Lyrics.
\newblock In \emph{ICCC}, 87--94.

\bibitem[{Neves, Fornari, and Florindo(2022)}]{neves2022generating}
Neves, P.; Fornari, J.; and Florindo, J. 2022.
\newblock Generating music with sentiment using Transformer-GANs.
\newblock In \emph{ISMIR}, 717--725.

\bibitem[{Nichols(2009)}]{nichols2009lyric}
Nichols, E. 2009.
\newblock Lyric-based rhythm suggestion.
\newblock In \emph{ICMC}.

\bibitem[{Payne(2019)}]{payne2019musenet}
Payne, C. 2019.
\newblock MuseNet.
\newblock \url{https://openai.com/blog/musenet}.

\bibitem[{Roberts et~al.(2018)Roberts, Engel, Raffel, Hawthorne, and
  Eck}]{roberts2018hierarchical}
Roberts, A.; Engel, J.; Raffel, C.; Hawthorne, C.; and Eck, D. 2018.
\newblock A hierarchical latent vector model for learning long-term structure
  in music.
\newblock In \emph{International conference on machine learning}, 4364--4373.
  PMLR.

\bibitem[{Sarmento et~al.(2023)Sarmento, Kumar, Chen, Carr, Zukowski, and
  Barthet}]{sarmento2023gtr}
Sarmento, P.; Kumar, A.; Chen, Y.-H.; Carr, C.; Zukowski, Z.; and Barthet, M.
  2023.
\newblock GTR-CTRL: instrument and genre conditioning for guitar-focused music
  generation with transformers.
\newblock In \emph{International Conference on Computational Intelligence in
  Music, Sound, Art and Design (Part of EvoStar)}, 260--275. Springer.

\bibitem[{Sheng et~al.(2021)Sheng, Song, Tan, Ren, Ye, Zhang, and
  Qin}]{sheng2021songmass}
Sheng, Z.; Song, K.; Tan, X.; Ren, Y.; Ye, W.; Zhang, S.; and Qin, T. 2021.
\newblock Songmass: Automatic song writing with pre-training and alignment
  constraint.
\newblock In \emph{Proceedings of the AAAI Conference on Artificial
  Intelligence}, volume~35, 13798--13805.

\bibitem[{Srivastava et~al.(2022)Srivastava, Duan, Shah, Wu, Tang, Li, and
  Yu}]{srivastava2022melody}
Srivastava, A.; Duan, W.; Shah, R.~R.; Wu, J.; Tang, S.; Li, W.; and Yu, Y.
  2022.
\newblock Melody generation from lyrics using three branch conditional
  LSTM-GAN.
\newblock In \emph{International Conference on Multimedia Modeling}, 569--581.
  Springer.

\bibitem[{Tan and Herremans(2020)}]{tan20music}
Tan, H.~H.; and Herremans, D. 2020.
\newblock Music FaderNets: Controllable Music Generation Based On High-Level
  Features via Low-Level Feature Modelling.
\newblock In \emph{ISMIR}, 109--116.

\bibitem[{Tian et~al.(2023)Tian, Narayan-Chen, Oraby, Cervone, Sigurdsson, Tao,
  Zhao, Chen, Chung, Huang et~al.}]{tian2023unsupervised}
Tian, Y.; Narayan-Chen, A.; Oraby, S.; Cervone, A.; Sigurdsson, G.; Tao, C.;
  Zhao, W.; Chen, Y.; Chung, T.; Huang, J.; et~al. 2023.
\newblock Unsupervised melody-to-lyric generation.
\newblock \emph{arXiv preprint arXiv:2305.19228}.

\bibitem[{Touvron et~al.(2023)Touvron, Martin, Stone, Albert, Almahairi,
  Babaei, Bashlykov, Batra, Bhargava, Bhosale et~al.}]{touvron2023llama}
Touvron, H.; Martin, L.; Stone, K.; Albert, P.; Almahairi, A.; Babaei, Y.;
  Bashlykov, N.; Batra, S.; Bhargava, P.; Bhosale, S.; et~al. 2023.
\newblock Llama 2: Open foundation and fine-tuned chat models.
\newblock \emph{arXiv preprint arXiv:2307.09288}.

\bibitem[{Vaswani et~al.(2017)Vaswani, Shazeer, Parmar, Uszkoreit, Jones,
  Gomez, Kaiser, and Polosukhin}]{vaswani2017attention}
Vaswani, A.; Shazeer, N.; Parmar, N.; Uszkoreit, J.; Jones, L.; Gomez, A.~N.;
  Kaiser, {\L}.; and Polosukhin, I. 2017.
\newblock Attention is all you need.
\newblock \emph{Advances in neural information processing systems}, 30.

\bibitem[{von R{\"u}tte et~al.(2023)von R{\"u}tte, Biggio, Kilcher, and
  Hofmann}]{von2023figaro}
von R{\"u}tte, D.; Biggio, L.; Kilcher, Y.; and Hofmann, T. 2023.
\newblock FIGARO: Controllable music generation using learned and expert
  features.
\newblock In \emph{The Eleventh International Conference on Learning
  Representations}.

\bibitem[{Wu and Yang(2023)}]{wu2023musemorphose}
Wu, S.-L.; and Yang, Y.-H. 2023.
\newblock MuseMorphose: Full-song and fine-grained piano music style transfer
  with one transformer VAE.
\newblock \emph{IEEE/ACM Transactions on Audio, Speech, and Language
  Processing}, 31: 1953--1967.

\bibitem[{Yang, Chou, and Yang(2017)}]{yang2017midinet}
Yang, L.-C.; Chou, S.-Y.; and Yang, Y.-H. 2017.
\newblock MidiNet: A convolutional generative adversarial network for
  symbolic-domain music generation.
\newblock In \emph{ISMIR}, 324--331.

\bibitem[{Yu, Srivastava, and Canales(2021)}]{yu2021conditional}
Yu, Y.; Srivastava, A.; and Canales, S. 2021.
\newblock Conditional LSTM-GAN for melody generation from lyrics.
\newblock \emph{ACM Transactions on Multimedia Computing, Communications, and
  Applications (TOMM)}, 17(1): 1--20.

\bibitem[{Zhang et~al.(2022)Zhang, Chang, Wu, Tan, Qin, Liu, and
  Zhang}]{zhang2022relyme}
Zhang, C.; Chang, L.; Wu, S.; Tan, X.; Qin, T.; Liu, T.-Y.; and Zhang, K. 2022.
\newblock Relyme: improving lyric-to-melody generation by incorporating
  lyric-melody relationships.
\newblock In \emph{Proceedings of the 30th ACM International Conference on
  Multimedia}, 1047--1056.

\bibitem[{Zhang, Yu, and Takasu(2023)}]{zhang2023controllable}
Zhang, Z.; Yu, Y.; and Takasu, A. 2023.
\newblock Controllable lyrics-to-melody generation.
\newblock \emph{Neural Computing and Applications}, 35(27): 19805--19819.

\end{thebibliography}

\end{document}